\documentclass[a4paper,11pt]{article}
\usepackage{pos}
\usepackage{combelow}

\DeclareMathOperator{\Tr}{Tr}

\title{Spatially inhomogeneous confinement-deconfinement phase transition in rotating QGP}
\author[a]{V.~V.~Braguta} 
\emailAdd{vvbraguta@theor.jinr.ru}
\author[b,c]{M.~N.~Chernodub}
\emailAdd{maxim.chernodub@univ-tours.fr}
\author[d]{Ya.~A.~Gershtein}
\emailAdd{gershtein.iaa@phystech.edu}
\author*[a]{A.~A.~Roenko}
\emailAdd{roenko@theor.jinr.ru}
\affiliation[a]{Bogoliubov Laboratory of Theoretical Physics, Joint Institute for Nuclear Research, \\ ul.~Joliot-Curie 6, Dubna, 141980, Russia}
\affiliation[b]{Institut Denis Poisson CNRS UMR 7013, Universit\'e de Tours, Universit\'e d'Orl\'eans, \\ Parc de Grandmont, Tours, 37200, France}
\affiliation[c]{Department of Physics, West University of Timi\cb{s}oara, \\ Bd.~Vasile P\^arvan 4, Timi\cb{s}oara, 300223, Romania}
\affiliation[d]{Moscow Institute of Physics and Technology, \\ Institutskii per.~9, Dolgoprudny, 141700, Russia}

\abstract{%
Using first-principles numerical simulations, we find a new spatially inhomogeneous phase in a rotating gluon plasma. This mixed phase simultaneously contains regions of both confining and deconfining states in thermal equilibrium, separated by a spatial transition.
The position of the boundary between the two phases is determined by the local critical temperature. We calculate the critical temperature of the local transition as a function of angular velocity and radius for a full (imaginary) rotating system and within a local thermalization approximation, and find an excellent agreement between these approaches.
An analytic continuation of the results to the domain of real angular frequencies indicates  that the confinement phase localizes at the periphery of the rotating system and the deconfinement phase appears closer to the rotation axis.
We argue that the anisotropy of the gluon action in the curved co-rotating background can quantitatively explain the remarkable property that the spatial structure of this inhomogeneous phase disobeys the picture based on a straightforward implementation of the Tolman-Ehrenfest law.
We also perform the first lattice simulation of rotating $N_f=2$ QCD which confirms that a similar picture is expected for theory with dynamical quarks. 
}

\FullConference{The 42nd International Symposium on Lattice Field Theory (LATTICE2025)\\
2-8 November 2025\\
Tata Institute of Fundamental Research, Mumbai, India\\}

\begin{document}
\maketitle

%%%%%%%%%%%%%%%%%%%%%%%%%%%%%%%%%%%%%%%%%%%%%%%%%%%%%%%%%%%%%%%%
\section{Introduction}\label{sec:Intro}

The fastest rotation in Nature is realized in heavy-ion collisions, which produce a highly vortical quark-gluon plasma (QGP)~\cite{Baznat:2013zx, Jiang:2016woz}.
The global polarization of $\Lambda$-, $\bar \Lambda$-hyperons measured by the STAR Collaboration indicates that the vorticity of the QGP in non-central collisions reaches the enormous value
$\Omega \sim 10^{22}$~Hz~$\sim 10$~MeV~\cite{STAR:2017ckg}.
At the characteristic droplet radius $r\sim ({\rm a\, few})\times\,{\rm fm}$ the velocity can reach $v \sim \Omega r \sim (0.1 \dots 0.3) c$.

Despite the fact that the vorticity is much smaller than the energy scale of Quantum Chromodynamics (QCD), $\Omega \ll \Lambda_{\rm QCD}$, the rotation may affect the properties of quark-gluon matter.
Lattice simulations are widely used to study the phase diagram, thermodynamic properties, and other aspects of rotating QCD~\cite{Yamamoto:2013zwa, Braguta:2020biu, Braguta:2021jgn, Braguta:2022str, Yang:2023vsw, Braguta:2023kwl, Braguta:2023yjn, Braguta:2023tqz, Braguta:2023iyx, Braguta:2024zpi, Braguta:2025yud}.
It was recently suggested that rotation could lead to a new mixed inhomogeneous phase in QCD~\cite{Chernodub:2020qah}.
In this state, the central and peripheral regions coexist in different phases separated by a spatial transition.
Lattice studies show that in rotating gluodynamics the confinement (deconfinement) phase is localized in the central (peripheral) part of the system~\cite{Braguta:2023iyx, Braguta:2024zpi}.
Note that this picture is qualitatively consistent with some effective models~\cite{Chen:2024tkr, Jiang:2024zsw}, whereas other approaches predict opposite phase arrangements~\cite{Chernodub:2020qah, Braga:2023qej}.

In this paper, we briefly summarize the results of Refs.~\cite{Braguta:2023iyx, Braguta:2024zpi}. We discuss the lattice formulation of the theory in the rotating frame, the mixed phase in rotating gluodynamics, the analytic continuation of the results, and the local thermalization approximation.
In the end, we present first results for mixed phase in QCD with dynamical quarks.

%%%%%%%%%%%%%%%%%%%%%%%%%%%%%%%%%%%%%%%%%%%%%%%%%%%%%%%%%%%%%%%%
\section{Lattice formulation of rotating QCD}\label{sec:Lattice}

It is convenient to describe the rotating system in a reference frame corotating together with the system around the $z$-axis.
Coordinates in this frame,
$x^\mu = (t, x, y, z) = (t, r \cos \varphi, r\sin\varphi, z)$,
are related to the coordinates in the laboratory frame,
$x^\mu_{\rm lab} = (t_{\rm lab}, r_{\rm lab} \cos \varphi_{\rm lab}, r_{\rm lab}\sin\varphi_{\rm lab}, z_{\rm lab})$,
by $\varphi = \varphi_{\rm lab} - \Omega t$, $r=r_{\rm lab}$, $z=z_{\rm lab}$, $t=t_{\rm lab}$.
This transformation induces the metric
\begin{equation}\label{eq:interval_Mink}
    ds^2 = g_{\mu\nu} dx^\mu dx^\nu = (1 - r^2 \Omega^2) dt^2 + 2y\Omega\, dt\,dx - 2x\Omega\, dt\,dy - dx^2 - dy^2 - dz^2\,,
\end{equation}
that encodes the effects of rotation. The Lagrangian of quark and gluon fields in the curved background~\eqref{eq:interval_Mink} can be easily written as follows~\cite{Braguta:2024zpi}
\begin{subequations}\label{eq:L}
\begin{align}
    \label{eq:L_psi}
    \mathcal{L}_\psi & = \bar\psi \left( i \gamma^\mu (D_\mu + \Gamma_\mu) - m \right) \psi
    = {\cal L}^{(0)}_{\psi} + {\cal L}^{(1)}_{\psi}
    \,, \\
    \label{eq:L_G}
    \mathcal{L}_G & = - \frac{1}{4 g_{YM}^2} g^{\mu \nu} g^{\alpha \beta} F_{\mu \alpha}^a F_{\nu \beta}^a
    = {\cal L}^{(0)}_{G} + {{\cal L}^{(1)}_{G}} + {{\cal L}^{(2)}_{G}}
    \,,
\end{align}
\end{subequations}
where $\mathcal{L}^{(n)}_{\psi,G} \propto \Omega^n$.
The quark Lagrangian is linear in the angular frequency $\Omega$:
$\mathcal{L}_\psi^{(0)}$ has the standard form for Dirac fermions,
and $\mathcal{L}_\psi^{(1)} = \Omega\cdot J_\psi^z$, where $J_\psi^z$ is the local angular momentum of the fermion field $\psi$.
On the contrary, the gluon Lagrangian has a part quadratic in $\Omega$:
the first contribution $\mathcal{L}_G^{(0)}$ is given by the Yang-Mills Lagrangian in the inertial laboratory frame,
and the second term has the standard appearance $\mathcal{L}_G^{(1)} = \Omega\cdot J_G^z$, where $J_G^z$ is the gluonic angular momentum density taken in the non-rotating limit,
while $\mathcal{L}_G^{(2)}$ is quadratic in $\Omega$ and contains only the squares of chromomagnetic fields.
However, the Hamiltonian of both rotating gluons and quarks,
\begin{equation} \label{eq:H_G_psi}
    {\hat H}_{\psi,G} = {\hat H}_{\psi,G}^{\rm lab} - {\Omega}\cdot {\hat{J}_{\psi,G}^z}\,,
\end{equation}
has the usual linear form, as it was explicitly shown in Refs.~\cite{Yang:2023vsw, Wang:2025mmv}.

The partition function of rotating QCD can be written using the path-integral representation,
\begin{equation}
    \label{eq:Partition}
    \mathcal{Z} = \Tr \left[e^{-\hat H/T}\right] =\Tr \left[ e^{-(\hat H^{\rm lab} - \Omega\cdot J^z)/T}\right] = \int \!D A\,D\psi\,D\bar\psi\, e^{-S[A,\psi,\bar\psi]}\,,
\end{equation}
where $S = S_G + S_\psi$ is the Euclidean action~\cite{Yamamoto:2013zwa}, and the inverse temperature~$1/T$ sets the system size in a compact Euclidean time, $\tau = i t$. Let's discuss the main features of the rotating QCD action:
\begin{enumerate}
    \item[a)] It is a complex-valued function for real $\Omega$ that leads to the sign problem.
    To overcome it, we perform the simulations at imaginary angular velocity, $\Omega_I = \partial \varphi_{\rm lab}/\partial\tau = - i \partial \varphi_{\rm lab}/\partial t = -i \Omega$, and  then analytically continue the results to the domain of $\Omega^2 > 0$.
    \item[b)] The Euclidean action is inhomogeneous and has the following structure:
    \begin{equation}\label{eq:S_structure}
        S = S_0 + \Omega_I S_1 + \Omega_I^2 S_2\,,
    \end{equation}
    where operators $S_1$, $S_2$ are constructed from the fields with coefficients which explicitly depend on transversal $x$-,$y$-coordinates
    (see Refs.~\cite{Yamamoto:2013zwa, Braguta:2021jgn, Braguta:2024zpi} for explicit expressions).
\end{enumerate}
Note that analytic continuation is allowed only for the bounded system that respects the causality condition, i.e. for any point of the system, the condition $\Omega r < 1$ must be satisfied.

The rotating gluodynamics alone has all these features. The sign problem originates from the terms linear in $\Omega$, which are common for both quark and gluon sectors of QCD, see Eqs.~\eqref{eq:L}.
Rotation also leads to asymmetry between chromoelectric and chromomagnetic fields via the quadratic term $S_2$, which may significantly affect the properties of the system.
To elaborate on the procedure of analytic continuation in an inhomogeneous system, we first consider the rotating gluons.

For rotating gluodynamics, we use the lattice setup of Refs.~\cite{Braguta:2023yjn,Braguta:2023kwl,Braguta:2023tqz,Braguta:2023iyx,Braguta:2024zpi}.
The homogeneous part $S_0$ is parametrized by the tree-level improved Symanzik gluon action,
while $S_1$ and $S_2$ are discretized using combinations of clovers and chair loops.
The simulations are performed on lattices of size $N_t \times N_z \times N_s^2$, with odd $N_s = N_x = N_y$.
In the directions $z$ and $t$, the standard periodic boundary conditions are implemented, whereas in the transversal directions $x$ and $y$ we use periodic (PBC) and open (OBC) boundary conditions to disentangle boundary and rotational effects.
The causality requirement  restricts the boundary velocity $v_I = \Omega_I R$, where $R = (N_s - 1)a/2$ is the characteristic transverse size of the system.
For the square lattice geometry, the causality requires $v_I < 1/\sqrt{2}$.

%%%%%%%%%%%%%%%%%%%%%%%%%%%%%%%%%%%%%%%%%%%%%%%%%%%%%%%%%%%%%%%%
\section{Mixed phase in rotating gluodynamics}\label{sec:MixedGluo}

\begin{figure}[t]
\centering
\includegraphics[width=1.0\textwidth]{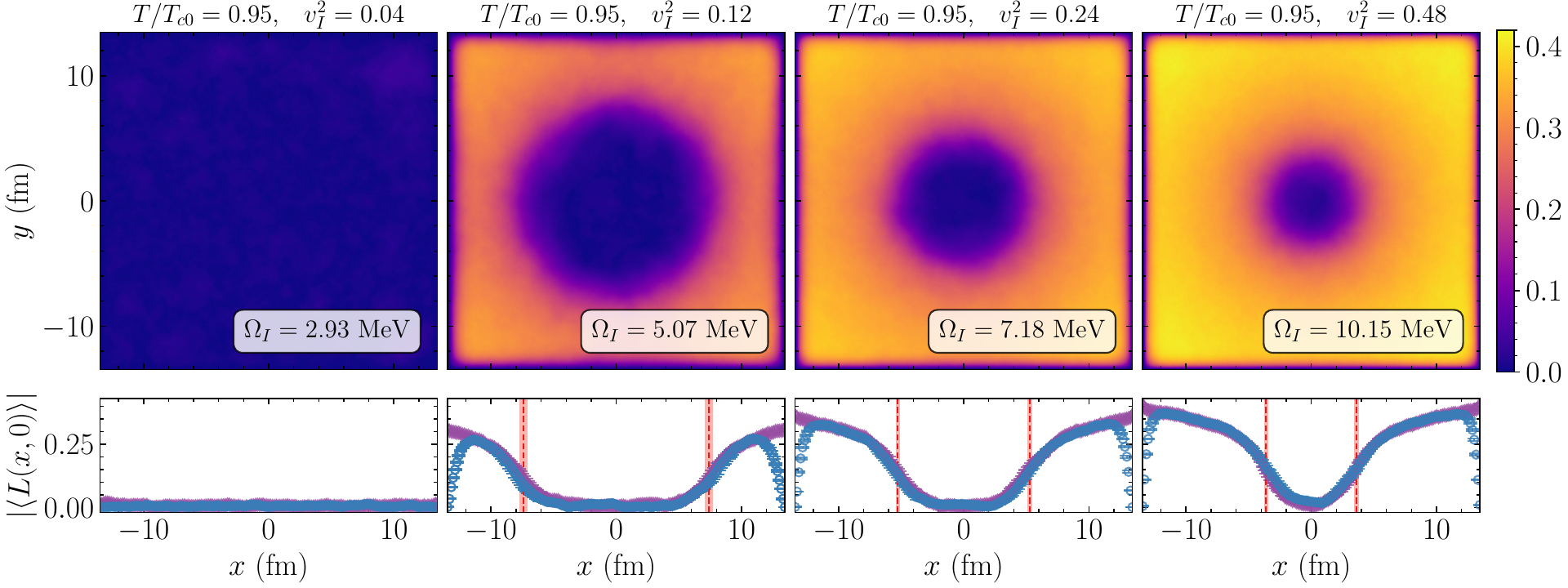}
\caption{%
(top) The distribution of the local Polyakov loop in the $x,y$-plane at the temperature $T/T_{c0} = 0.95$ and several values of the angular velocity (also shown in units of $v_I = \Omega_I R$ with $R=13.5$~fm) for a lattice of size $5\times 30\times 181^2$ with open boundary conditions. 
(bottom) The Polyakov loop at the $x$-axis. The vertical lines mark the phase boundaries with shaded uncertainties.
The violet (blue) data points correspond to periodic (open) boundary conditions.
}
\label{fig:PLxy_gluo}
\end{figure}

The Polyakov loop plays the role of the order parameter that is commonly used to distinguish the low-temperature confinement and high-temperature deconfinement phases.
To study their spatial structure, we consider the local Polyakov loop in the transversal plane,
\begin{equation}
    \label{eq:P_loop}
    L(x,y) = \frac{1}{N_z} \sum_{z=0}^{N_z-1} L(x,y,z)\,, \qquad
    L(x,y,z) = \Tr \left[ \prod_{\tau = 0}^{N_t - 1} U_4(\tau, x,y,z) \right]\,,
\end{equation}
where $U_4(\tau, x,y,z)$ denotes a temporal link.
We measure $L(x,y)$ for different lattice parameters and find that two phases may coexist for $T\lesssim T_{c0}$, where $T_{c0}$ is the critical temperature of the non-rotating system.
Figure~\ref{fig:PLxy_gluo} shows the spatial distribution of the local Polyakov loop at $T/T_{c0} = 0.95$ for several values of (imaginary) angular velocity. 
At vanishing rotation, the local Polyakov loop is approximately zero in the whole system, indicating confinement.
As the (imaginary) angular velocity increases, the periphery of the system goes to the deconfinement phase with a non-zero Polyakov loop, and the radius of the inner confinement region shrinks.
This picture also persists for the largest velocity, $v_I^2 \simeq 0.5$, when corners of the system move with almost the speed of light.
Note that the results obtained with PBC and OBC in the transversal directions are consistent in the bulk, indicating the independence of the bulk properties on the conditions on the spatial boundaries of the system. The rotational symmetry is also restored since the boundary between phases has approximately a circular form.
%%%%%%%%%%%%%%%%%%%%%%%%%%%%%%%%
%%%%%%%%%%%%%%%%%%%%%%%%%%%%%%%%

To parametrize the behavior of the phase boundary, we introduce a critical temperature of the local transition, $T_c(r)$. 
It is defined as the temperature of the system\footnote{It should not be confused with the \textit{local equilibrium temperature} $T(r)$, which is given by the Tolman-Ehrenfest law.},  when the spatial transition occurs at the radius $r$.
We determine $T_c(r)$ from the peak of the local susceptibility of the Polyakov loop, $\chi(r) = \langle |L(r)|^2\rangle - \langle |L(r)|\rangle^2$, fitted by a Gaussian function. 
To reduce the statistical error, we calculate it within a thin cylinder of thickness $\delta r$. In addition, to suppress boundary effects, we discard $\delta b$ layers adjacent to the boundary.
We found that the finiteness of $\delta r$ and $\delta b$ brings only a minor systematic error to the results.

%%%%%%%%%%%%%%%%%%%%%%%%%%%%%%%%
\begin{figure}[t]
\centering
\includegraphics[width=0.49\textwidth]{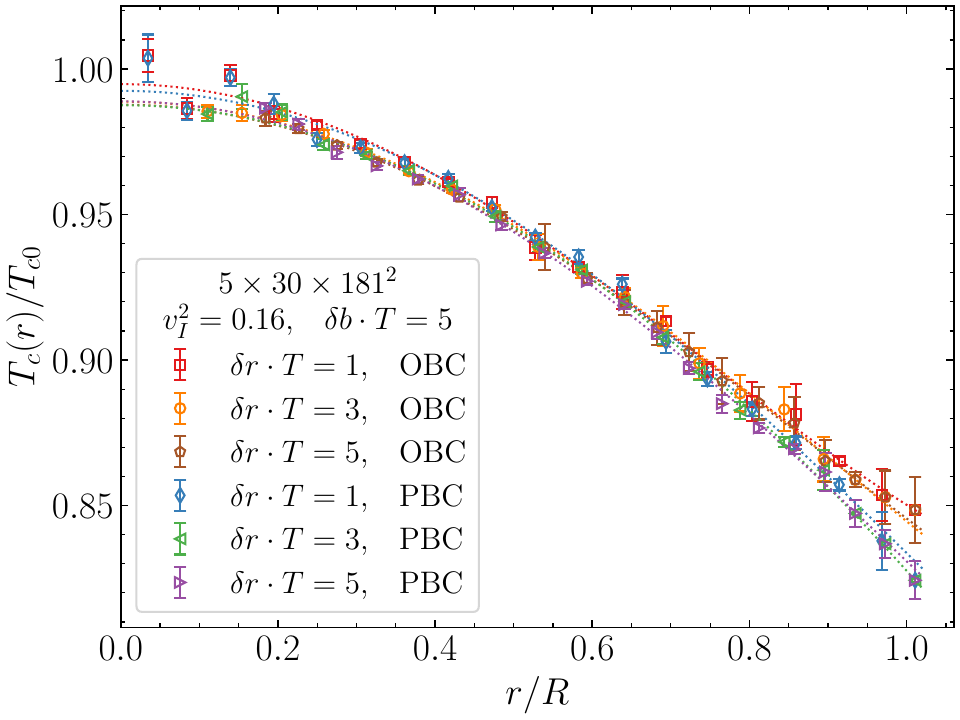}
\hfill
\includegraphics[width=0.49\textwidth]{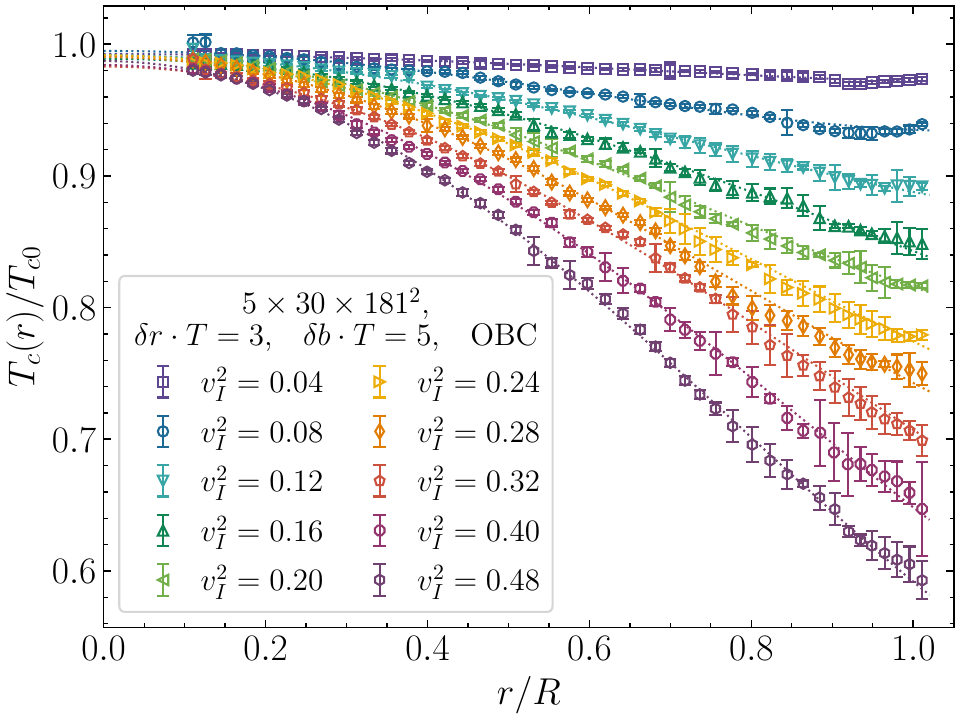}
\caption{%
The critical temperature $T_c(r)$ of the local transition as a function of radius $r$ at $v_I^2 = 0.16$ for open and periodic boundary conditions (left) and at various velocities $v_I$ for open boundary conditions (right). 
}
\label{fig:Tc_vs_r}
\end{figure}
%%%%%%%%%%%%%%%%%%%%%%%%%%%%%%%%
The critical temperature $T_c(r)$ as a function of radius $r$ is shown in Fig.~\ref{fig:Tc_vs_r} at various velocities for the lattice of size $5\times 30\times 181^2$.
Similar results were obtained on lattices with different $N_t = 4, \, 6$ and the same aspect ratios $N_z/N_t$, $(N_s-1)/N_t$.
We fit all sets of the data with the even power polynomial of $r/R$ up to a quartic term, and conclude that the critical temperature decreases with imaginary angular velocity as follows~\cite{Braguta:2024zpi}:
\begin{equation}\label{eq:Tc_r}
    \frac{T_c(r)}{T_{c0}}
    = 1 - (\Omega_I r)^2 \left( \kappa_2 - \kappa_4 \left( \frac{r}{R} \right)^2\right)\,.
\end{equation}
The quadratic coefficient $\kappa_2$ is universal, i.e., it doesn't depend on the type of boundary conditions and the transverse lattice size. On the contrary, the quartic coefficient $\kappa_4$ depends on these lattice parameters~\cite{Braguta:2024zpi}. Using the data in the bulk, we get $\kappa_2 = 0.902(33)$ in the continuum limit.

%%%%%%%%%%%%%%%%%%%%%%%%%%%%%%%%%%%%%%%%%%%%%%%%%%%%%%%%%%%%%%%%
\section{Action decomposition and analytic continuation}\label{sec:Decomposition}

The procedure of the analytic continuation may seem non-trivial for an inhomogeneous system.
To clarify it and investigate the origin of this peculiar behavior of rotating gluon plasma, we rewrite the action~\eqref{eq:S_structure} in the following form,
\begin{equation}
    \label{eq:S_lambdas}
    S = S_0 + \lambda_1 \Omega_I S_1 + \lambda_2 \Omega_I^2 S_2\,,
\end{equation}
where the factors $\lambda_1$, $\lambda_2$ are introduced. These factors allow us to switch on/off the operators $S_1$ and $S_2$ in the simulations. Besides the physical regime Im12 ($\lambda_1 =\lambda_2 = 1$), the case $\lambda_1 = 0$ is of particular interest.
There is no sign problem in this case while the system remains inhomogeneous. In this setup, the results of simulation for regime Re2 ($\lambda_1 = 0$, $\lambda_2 \Omega_I^2 = - \Omega_I^2 < 0$, or $\Omega^2 > 0$) are expected to be related to those for regime Im2 ($\lambda_1 = 0$, $\lambda_2 \Omega_I^2 = \Omega_I^2 > 0$) by analytic continuation.

%%%%%%%%%%%%%%%%%%%%%%%%%%%%%%%%
\begin{figure}[th]
\centering
\includegraphics[width=0.49\textwidth]{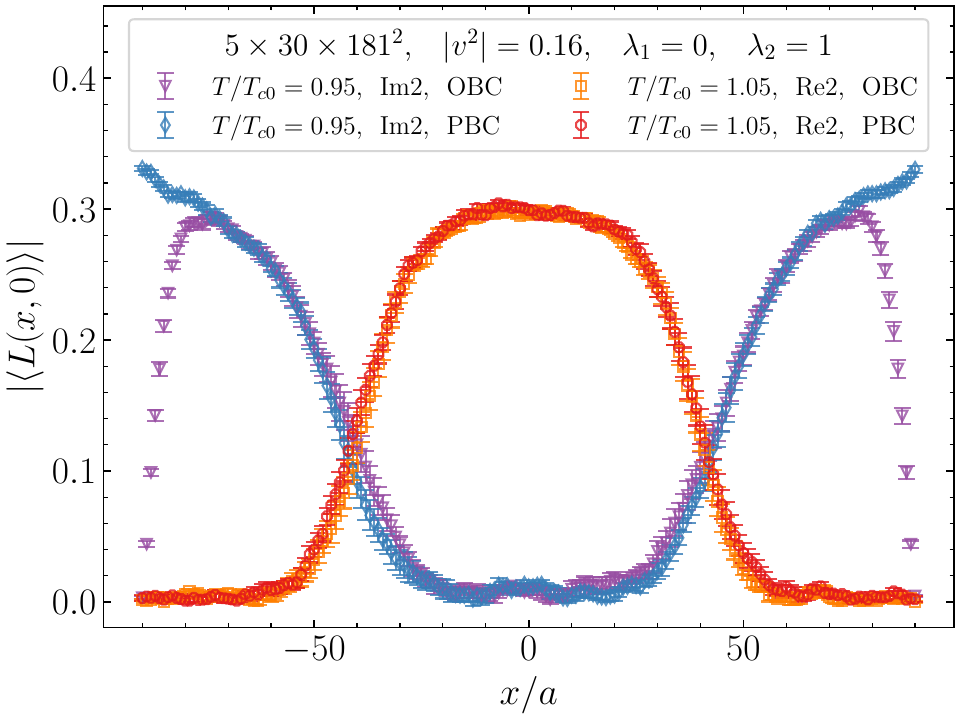}
\hfill
\includegraphics[width=0.49\textwidth]{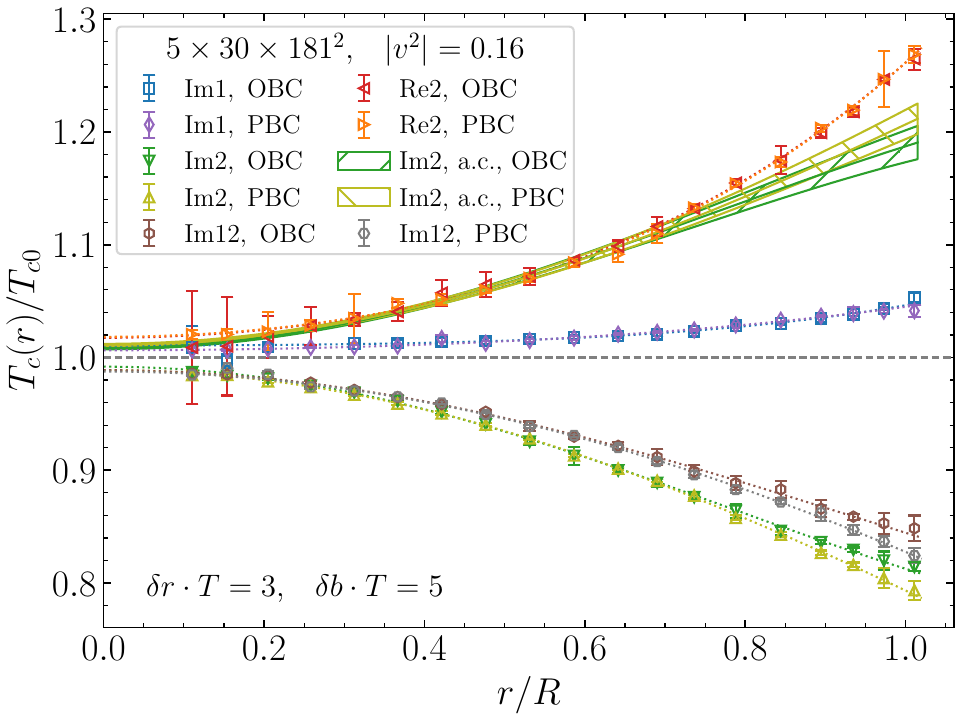}
\caption{%
(left) The local Polyakov loop as a function of $x$ coordinate for lattice of size $5\times 30\times 181^2$ with OBC/PBC for the regimes Im2 and Re2. Data are shown, respectively, for $|v^2| =  0.16$ at $T/T_{c0} = 0.95$ with imaginary angular velocity (Im2) and at $T/T_{c0} = 1.05$ with real angular velocity (Re2).
(right) The critical temperature $T_c(r)$ of the local transition as a function of radius for different regimes of rotation.
Hatched bands represent the analytic continuation of the Im2 results to the domain of real $\Omega$ using Eq.~\eqref{eq:Tc_r}.
}
\label{fig:Im2_vs_Re2}
\end{figure}
%%%%%%%%%%%%%%%%%%%%%%%%%%%%%%%%
The spatial profile of the local Polyakov loop for imaginary and real values of the parameter $\Omega$ (i.e. for regimes Im2 and Re2, respectively) is shown in Fig.~\ref{fig:Im2_vs_Re2}~(left).
The results are presented for $\Omega^2 < 0$ at temperature $T = T_{c0} - \Delta T$ (regime Im2) and for $\Omega^2 > 0$ at $T = T_{c0} + \Delta T$ (regime Re2), with $\Delta T = 0.05 T_{c0}$ and $|v^2| = |(\Omega R)^2| = 0.16$.
One can see that the boundary between phases is located at the same radius in both cases, but the arrangement of phases is reversed.
Note that the spatial boundary conditions affect the results only in the vicinity of the system edge.

The radial dependence of the critical temperature $T_c(r)$ is shown in Fig.~\ref{fig:Im2_vs_Re2}~(right) for different regimes.
These results indicate that, within a statistical uncertainty,
the critical temperatures for regimes Im2 and Re2 are connected through the analytical continuation procedure $\Omega^2 \leftrightarrow -\Omega_I^2$ in Eq.~\eqref{eq:Tc_r}, although a minor discrepancy appears in the vicinity of the boundary.\footnote{This minor discrepancy may indicate that the next-order terms in the fitting function may be required.}
It is also seen that the results for the Im2-regime are quite close to the physical regime Im12 at imaginary rotation.
These two regimes, Im2 and Im12, differ by the accounting of the linear term $S_1$.
Effects of this term may be considered separately in the additional regime Im1 ($\lambda_1 = 1$, $\lambda_2 = 0$).
The critical temperature increases in this regime, in contrast to Im2 and Im12, but the behavior of $T_c(r)$ with radius turns out to be very weak.
So, the linear term $S_1$ and the quadratic term $S_2$ produce different effects on the critical temperature, with the dominance of the quadratic one, which generates asymmetry between chromoelectric and chromomagnetic fields.
Note that this result resembles the decomposition of the moment of inertia into mechanical and magnetic parts (see Ref.~\cite{Braguta:2023tqz}), where the magnetic contribution, associated with $S_2$, dominates at temperatures near $T_{c0}$.

%%%%%%%%%%%%%%%%%%%%%%%%%%%%%%%%
\begin{figure}[th]
\centering
\includegraphics[width=0.55\textwidth]{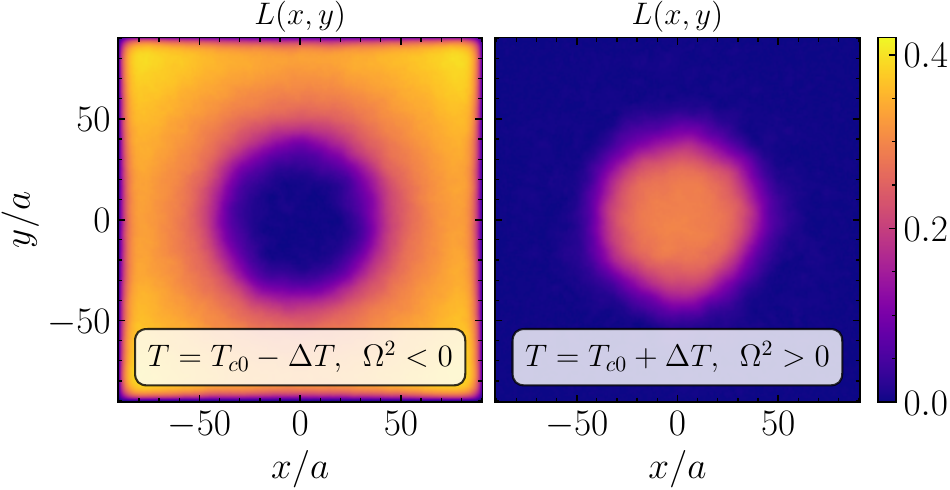}
\caption{%
The mixed inhomogeneous phase for imaginary and real rotations.
}
\label{fig:PLxy_qual}
\end{figure}
%%%%%%%%%%%%%%%%%%%%%%%%%%%%%%%%
To conclude this section, we show a schematic representation of the mixed inhomogeneous phase for imaginary and real rotating systems in Fig.~\ref{fig:PLxy_qual} at slow rotation, when only the quadratic in $\Omega$ term in Eq.~\eqref{eq:Tc_r}, is sufficient to describe the critical temperature and its analytic continuation.
This phase arrangement contradicts the expectation based on the Tolman-Ehrenfest law and corresponding local equilibrium temperature  (see discussion in Ref.~\cite{Braguta:2024zpi}).
By contrast, a lattice study of gluodynamics with different gravitational background~--- in an accelerated frame~--- show an agreement with the TE law prediction due to a specific transformational symmetry of the action~\cite{Braguta:2026nfy}.

%%%%%%%%%%%%%%%%%%%%%%%%%%%%%%%%%%%%%%%%%%%%%%%%%%%%%%%%%%%%%%%%
\section{Approximation of local thermalization}\label{sec:Local}

The study of an inhomogeneous system is complicated by the boundary effects.
To disentangle them from the rotational effects, one can apply an approximation of local thermalization~\cite{Braguta:2024zpi}.
In this approach, we consider first a small spatial subsystem of a large rotating system.
One can take a vicinity of a point $(x_0, y_0) = (r_0, 0)$ at radial distance $r_0$ from the axis. Then, we set the coefficients in the action to constant values taken at this point.
As a result, we arrive to the homogeneous gluon system with the following anisotropic Euclidean action:
\begin{multline}\label{eq:S_local}
	S_{G} = \frac{1}{2 g_{YM}^{2}} \int\! d^{4}x \ \Big[
    F^a_{x \tau} F^a_{x \tau} + F^a_{y \tau} F^a_{y \tau} + F^a_{z \tau} F^a_{z \tau} 
    + F^a_{x z} F^a_{x z} + {}\\
    + \left(1 + u_I^2\right) F^a_{y z} F^a_{y z} 
    + \left(1 + u_I^2\right) F^a_{x y} F^a_{x y}
    - 2 u_I \left(F^a_{y x} F^a_{x \tau} + F^a_{y z} F^a_{z \tau}\right)
    \Big]\, ,
\end{multline}
where $u_I = \Omega_I r_0$ is a coordinate-independent parameter. In this approach, which we call 
the approximation of local thermalization, observables depend on the (imaginary) velocity parameter $u_I$, and not on the angular velocity or radius itself, as expected from the leading term in Eq.~\eqref{eq:Tc_r}.

\begin{figure}[t]
\centering
\includegraphics[width=0.49\textwidth]{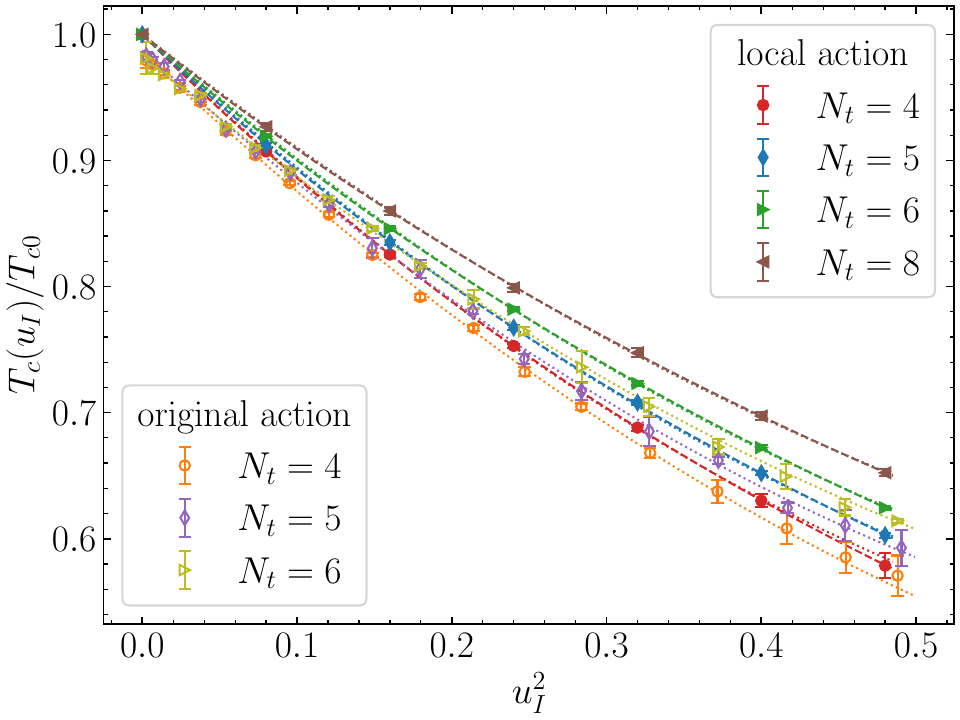}
\hfill
\includegraphics[width=0.49\textwidth]{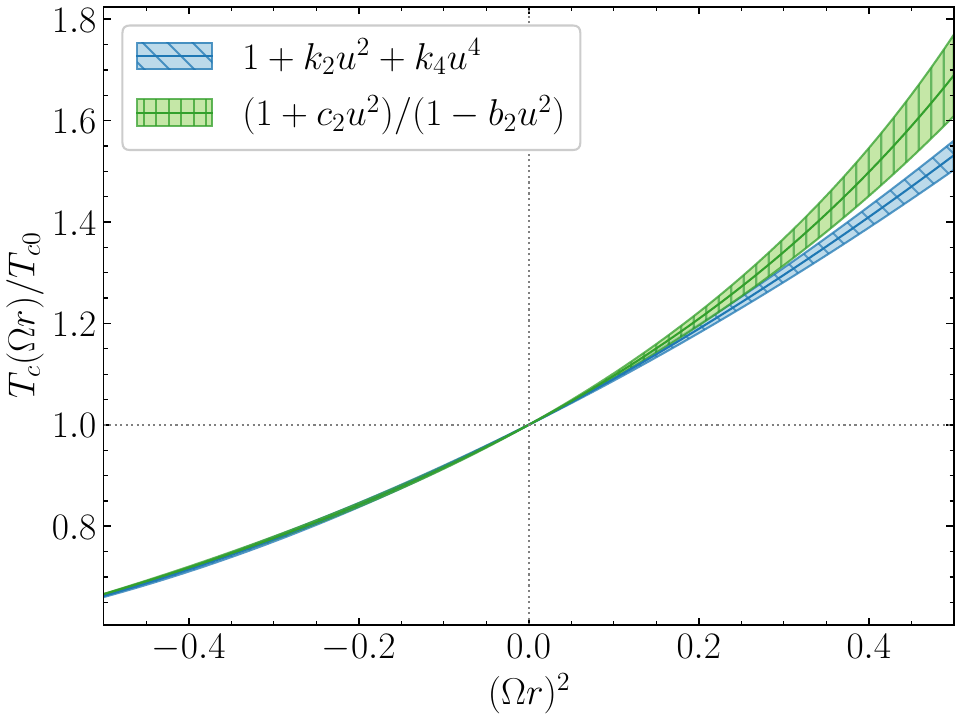}
\caption{%
(left) The critical temperature in the purely gluon system with local action~\eqref{eq:S_local} (filled points) as a function of $u_I$.
The dotted (dashed) lines represent the best fit of the data by the polynomial (rational) function~\eqref{eq:fit_u}.
In addition, we show the critical temperatures calculated for the rotating system with the original action~\eqref{eq:S_structure} (empty points) at $v_I = 0.48$.
(right) The fitting functions~\eqref{eq:fit_u} in the continuum limit.
}
\label{fig:Tc_vs_u_local}
\end{figure}

We discretize the action~\eqref{eq:S_local} in the same way as for the full rotating system studied in Sections~\ref{sec:MixedGluo},~\ref{sec:Decomposition}, and calculate the critical temperatures using the Polyakov loop susceptibility for different $u_I$ on lattices of size $N_t \times N_\sigma^3$ with $N_t = 4,\,5,\,6,\,8$ and $N_\sigma/N_t = 4$.
The results for the homogeneous but anisotropic system~\eqref{eq:S_local} are shown in Fig.~\ref{fig:Tc_vs_u_local} (left), as well as the rescaled critical temperature for the original rotating gluon system at $v_I^2 = 0.48$.
One observes a very good agreement between the two approaches: the local critical temperature of the original, anisotropic and inhomogeneous theory~\eqref{eq:S_structure} coincides with the local approximation given by the anisotropic but homogeneous action~\eqref{eq:S_local}, where the local anisotropy is extended to the whole bulk volume of the system. The critical temperature of the local system is well described by the polynomial and rational functions,
\begin{equation}\label{eq:fit_u}
    \frac{T_c(u)}{T_{c0}} = 1 + k_2 u^2 + k_4 u^4\,, \hspace{3em} \textrm{and} \hspace{3em}
    \frac{T_c(u)}{T_{c0}} = \frac{1 + c_2 u^2}{1 - b_2 u^2}\,,
\end{equation}
where the best-fit coefficients take the following values after extrapolation to the continuum limit:
$k_2 = 0.869(31)$, $k_4 = 0.388(53)$, and 
$c_2 = 0.206(66)$, $b_2 = 0.694(101)$.
These functions coincide at imaginary rotation, but they slightly differ in the domain of real $u=\Omega r_0$, thus providing us with an estimation of the systematic uncertainty of the analytic continuation (see Fig.~\ref{fig:Tc_vs_u_local}).

From the expression~\eqref{eq:S_local}, one can see how the curved spacetime metric~\eqref{eq:interval_Mink} modifies the gluon action.
The quadratic in $u_I^2$ terms, which give the dominant contribution, generate an asymmetry in effective coupling for two chromomagnetic components of the gluon field, somewhat similar to that considered in Ref.~\cite{Karsch:1982ve}.
This asymmetry affects the dynamics of the gluon field that cannot be accounted for by the TE law.

%%%%%%%%%%%%%%%%%%%%%%%%%%%%%%%%
%%%%%%%%%%%%%%%%%%%%%%%%%%%%%%%%

%%%%%%%%%%%%%%%%%%%%%%%%%%%%%%%%%%%%%%%%%%%%%%%%%%%%%%%%%%%%%%%%
\section{Mixed phase in rotating QCD}\label{sec:MixedQCD}

\begin{figure}[t]
\centering
\includegraphics[width=1.0\textwidth]{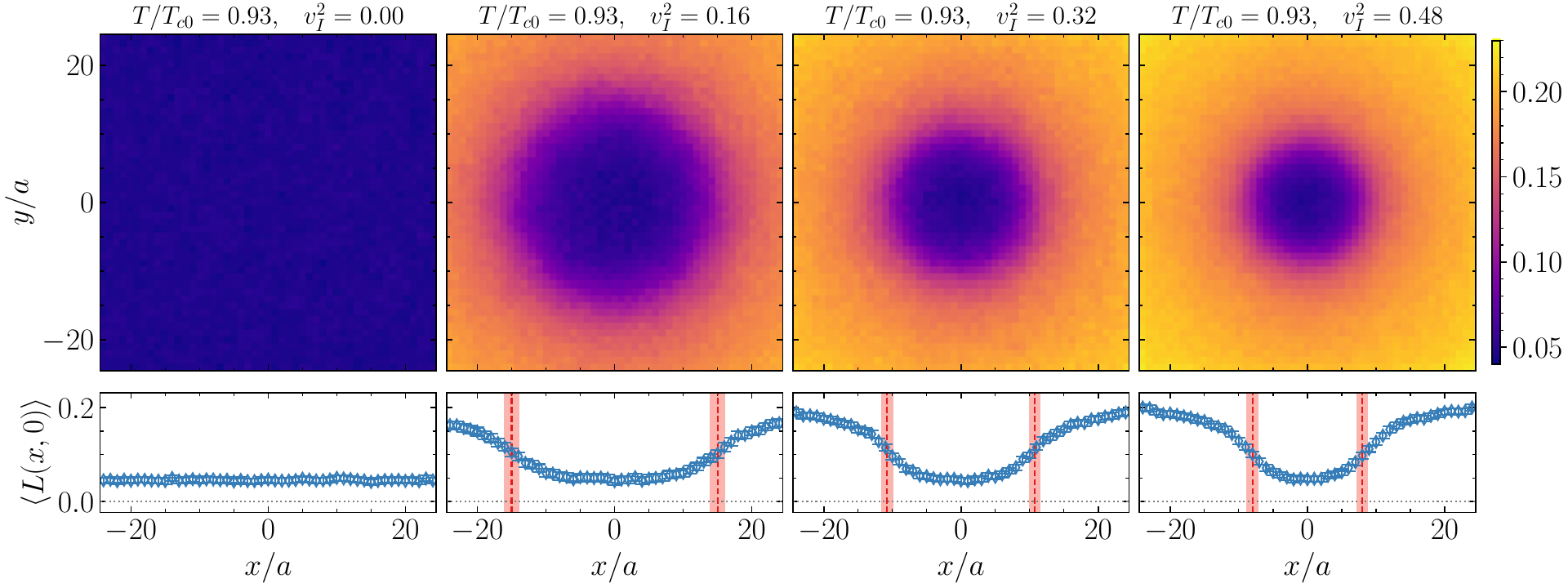}
\caption{%
The same as in Fig.~\ref{fig:PLxy_gluo} but for rotating QCD with $N_f=2$ dynamical quarks. The results are obtained on the lattice of size $4\times 24\times 49^2$ with PBC in the $x,y$-directions at the temperature $T/T_{c0} = 0.93$.
}
\label{fig:PLxy_qcd}
\end{figure}

A similar mixed-phase structure is expected in rotating QCD.
Indeed, quarks contribute only to the linear term $S_1$ of the action, which plays a sub-leading role~\cite{Braguta:2022str}.
Using the lattice setup of Ref.~\cite{Braguta:2022str} for rotating QCD with $N_f = 2$ dynamical clover-improved Wilson quarks, we calculated the distribution of the local Polyakov loop in the $x,y$-plane at the temperature $T/T_{c0} = 0.93$ for $m_{PS}/m_V = 0.80$ and several values of the rotational velocity (see Fig.~\ref{fig:PLxy_qcd}).
We found qualitatively the same pattern as in rotating gluodynamics. Note that quantitative features are expected to depend on the pion mass. Moreover, since QCD exhibits two overlapping, chiral and confinement crossover transitions, the rotation should also convert the chiral symmetry breaking-restoration into a spatial transition, which may be shifted with respect to the confinement-deconfinement transition~\cite{Sun:2023kuu,Singha:2024tpo}. This question will be considered in forthcoming studies.
%%%%%%%%%%%%%%%%%%%%%%%%%%%%%%%%
%%%%%%%%%%%%%%%%%%%%%%%%%%%%%%%%

%%%%%%%%%%%%%%%%%%%%%%%%%%%%%%%%%%%%%%%%%%%%%%%%%%%%%%%%%%%%%%%%
\section{Conclusions}\label{sec:Conclusions}

Using the first-principles lattice simulations, we studied a mixed inhomogeneous phase in rotating gluodynamics. In this state, the confinement and deconfinement regions coexist at thermal equilibrium and are separated by the spatial transition at a fixed distance from the rotation axis (see Fig.~\ref{fig:PLxy_qual}). The radial position of this transition~--- the boundary between phases~--- is described by the critical temperature, which is an increasing function of the (real) angular velocity and radius. This growth of the critical temperature cannot be accounted for by the Tolman-Ehrenfest law, which predicts an opposite behavior. We conclude that this unusual phase structure of rotating gluonic fields arises due to the spatial anisotropy of the gluon action in the background of an effective gravitational field that arises in the non-inertial frame corotating with the gluon matter. The same property is also observed for rotating QCD with quarks.

%%%%%%%%%%%%%%%%%%%%%%%%%%%%%%%%%%%%%%%%%%%%%%%%%%%%%%%%%%%%%%%%
\acknowledgments
The work of VVB, YaAG and AAR has been carried out using computing resources of the Federal collective usage center Complex for Simulation and Data Processing for Mega-science Facilities at NRC ``Kurchatov Institute'', http://ckp.nrcki.ru/, and
the heterogeneous computing platform HybriLIT (LIT, JINR).
The work of VVB and AAR, which consisted in the lattice calculation of the observables used in the paper, was supported by the Russian Science Foundation (project no. 23-12-00072). 
The work of MNC was funded by the EU’s NextGenerationEU instrument through the National Recovery and Resilience Plan of Romania - Pillar III-C9-I8, managed by the Ministry of Research, Innovation and Digitization, within the project entitled ``Facets of Rotating Quark-Gluon Plasma'' (FORQ), contract no.~760079/23.05.2023 code CF 103/15.11.2022.

%%%%%%%%%%%%%%%%%%%%%%%%%%%%%%%%%%%%%%%%%%%%%%%%%%%%%%%%%%%%%%%%
\bibliographystyle{JHEP}
\bibliography{plasma.bib}

\end{document}